\documentclass{elsart5p}
\usepackage{amsmath}
\usepackage{amssymb}
\usepackage{graphicx}
\usepackage{subfigure}
\usepackage{latexsym}
\usepackage{latexsym}
\usepackage[square,numbers,sort&compress,comma]{natbib}
\usepackage{mathptm} 
\usepackage{psfrag}

\begin{document}

\begin{frontmatter}

 \title{
        Modified Associate Formalism without Entropy Paradox:\\
        Part II. Comparison with Similar Models
        }

 \author{Dmitry N. Saulov\corauthref{cor}\thanksref{label1}}\ead{d.saulov@uq.edu.au},
 \quad
 \author{A. Y. Klimenko\thanksref{label1}}
 \corauth[cor]{
               Corresponding author:
               School of Engineering,
               The University of Queensland,
               St. Lucia, QLD 4072, Australia.
               Phone: (07)3365-3677
               Fax:  (07)3365-3670
              }

 \address[label1]{School of Engineering, The University of Queensland, AUSTRALIA}

\journal {Journal of Alloys and Compounds} \volume {473(1–2)} \pubyear{2009} \firstpage{157} \lastpage{162}

\begin{abstract}
The modified associate formalism is compared with similar models, such as the classical
associate model, the associate species model and the modified quasichemical model.
Advantages of the modified associate formalism are demonstrated.
\end{abstract}


\end{frontmatter}

\newcommand {\p}{q}
\newcommand{\kk}{{k_1,\dots,k_r}}
\newcommand{\jj}{{[j]}}
\newcommand{\mm}{{(m)}}
\newcommand{\set}{{\mathfrak{S}_m}}
\newcommand{\sett}{\mathfrak{S}^\circ_m}

\section{Introduction}
In this paper we compare the modified associate formalism (MAF) proposed in the first
paper~\cite{Saulov_MAF1} with the classical associate model (CAM) described by Prigogine
and Defay~\cite{PrigogineDefay} and the associate species model (ASM) described by
Besmann and Spear~\cite{Besmann02}. Where it is possible, we compare MAF with the
classical quasichemical model (CQM) of Fowler and Guggenheim~\cite{Fowler_Guggenheim}. We
also compare MAF with  two modifications of CQM. The first modification suggested by
Pelton and Blander \cite{Pelton_86} include:
\begin{itemize}
\item selection of the coordination numbers that allow one to set the
composition of maximum ordering so as to comply with the experimental data;
\item an empirical expansion of the molar
Gibbs energy of the quasichemical reaction as polynomials in terms of coordination
equivalent mole fractions of solution components, which is used for fitting available
experimental data.
\end{itemize}
The first version of the modified quasichemical model is referred to as MQM(1986) in the
present study. Further modifications of the quasichemical model introduced by Pelton et
al. \cite{Pelton_00} are:
\begin{itemize}
\item expansion of the Gibbs energy of the quasichemical reaction as polynomials in
pair fractions (instead of equivalent mole fractions of solution components), which
provides advantages in data-fitting;
\item dependence of coordination numbers of the solution components on the mole
numbers of different pairs that allow one to set the composition of maximum ordering so
as to comply with the experimental data for each binary system individually.
\end{itemize}
We refer to the second version of the modified quasichemical model as MQM(2000).

\section{Configurational Entropy of Mixing in Binary Solutions}
In this section, we compare the molar configurational entropies of mixing given by the
solution models mentioned above with theoretical boundaries for the configurational
entropy of mixing. These boundaries are discussed immediately below.

The configurational entropy of mixing is a measure of disorder and has its
maximum value for an ideal solution which is completely disordered. For any
solid or liquid solution $A-B$, the configurational entropy of mixing can not
exceed that of ideal solution. On the other hand, when vacancies are not taken
into account (which is the case in all considered models), the configurational
entropy of mixing can not be negative. The values of molar configurational
entropy of mixing which are outside of these boundaries are paradoxical.

For each model, we consider the following three limiting cases.
\begin{enumerate}
\item The case of ideal solution of $A$-particles and $B$-particles. In this case,
the Gibbs free energy changes on forming associates that consist of both $A$-particles
and $B$-particles. The configurational entropy of mixing should reduce to that of the
ideal solution model.
\item The case of immiscible components $A$ and $B$. In this case, the configurational entropy of
mixing should equals to zero.
\item The case of the highly ordered solution $A-B$. In this case, the Gibbs free energy change
on forming of the associates of a particular composition and of a particular spacial
arrangement of particles approaches $-\infty$. If the quasichemecal model is considered,
the Gibbs free energy change on forming $(A-B)$-bonds approaches $-\infty$. The
configurational entropy of mixing should be equal to zero at the composition of maximum
ordering.
\end{enumerate}

\subsection{Classical Associate Model}
Consider a solution of $x_A$ moles of the component $A$ and $x_B=1-x_A$ moles of the
component $B$.  The molar configurational entropy of mixing $s_\text{conf}$ given by the
classical associate model~\cite{PrigogineDefay} is expressed as
\begin{equation}\label{eq associated S_config}
s_\text{conf}=-R(n_{A_1}\ln(x_{A_1})+n_{B_1}\ln(x_{B_1})+\sum\limits_{i,j\geq
1}n_{A_{i}B_{j}}\ln(x_{A_{i}B_{j}})).
\end{equation}
Here, $n_{A_1}$, $n_{B_1}$ and $n_{A_{i}B_{j}}$  are the mole numbers (in one mole of the
solution $A-B$), while $x_{A_1}$, $x_{B_1}$ and $x_{A_{i}B_{j}}$ are the molar fractions
of $A_1$-monoparticles, $B_1$-monoparticles and ${A_{i}B_{j}}$-associates, respectively;
$R$ is the universal gas constant. Note that compositions of associates and sizes of
associates ($i+j$) are determined by a modeller. The equilibrium values of the mole
numbers are determined by minimisation of the Gibbs free energy of the solution subject
to the mass balance and nonnegativity constraints (see, for example, the monograph by
Prigogine and Defay~\cite{PrigogineDefay} for more details).

\begin{figure}
\begin{center}
\includegraphics[width=\columnwidth]{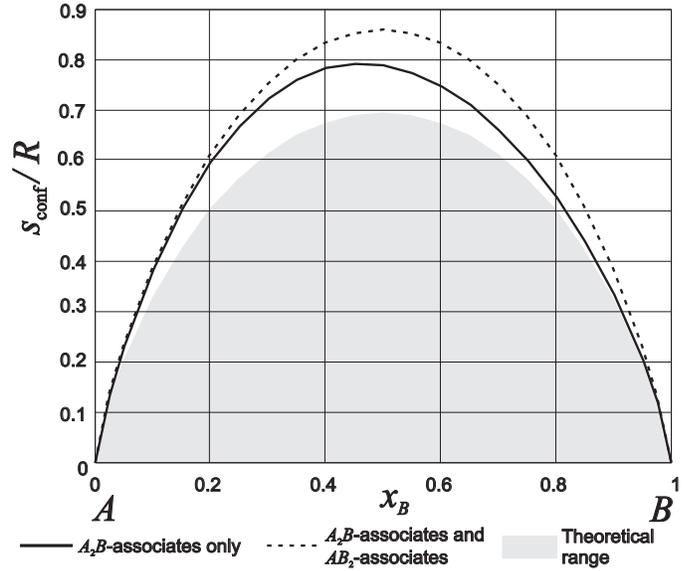}%
\caption {Molar configurational entropy of mixing given by the classical associate model
in the limit of ideal solution of $A$-particles and $B$-particles.}
\label{fig cam ideal}%
\end{center}
\end{figure}

Now consider the limit when the Gibbs free energy changes on forming the associates equal
to zero. As pointed out by L\"{u}ck et al.~\cite{Luck_entropy_paradox}, the classical
associate model gives paradoxical values of the molar configurational entropy of mixing,
which are larger than those given by the ideal solution model. Fig.~\ref{fig cam ideal},
for example, shows the configurational entropy curves given by the classical associate
model where only $A_2B$-associates are considered (solid line) and where
$A_2B$-associates and $AB_2$-associates are considered (dashed line).

As shown in the figure, the curves are outside of the theoretically possible range. Note
also that the entropy curve is not symmetric with respect to vertical line with $x_B=1/2$
when only $A_2B$-associates are considered. To obtain a symmetric entropy curve (which is
expected in the limit of ideal solution) one should consider both $A_2B$-associates and
$AB_2$-associates.

Consider the limit when the Gibbs free energy changes on forming all associates approach
$+\infty$. In this case, no associates form and the configurational entropy of mixing
equals to that of the ideal solution model. In theory, however, the configurational
entropy of mixing should be zero for any $x_B$.

\begin{figure}
\begin{center}
\includegraphics[width=\columnwidth]{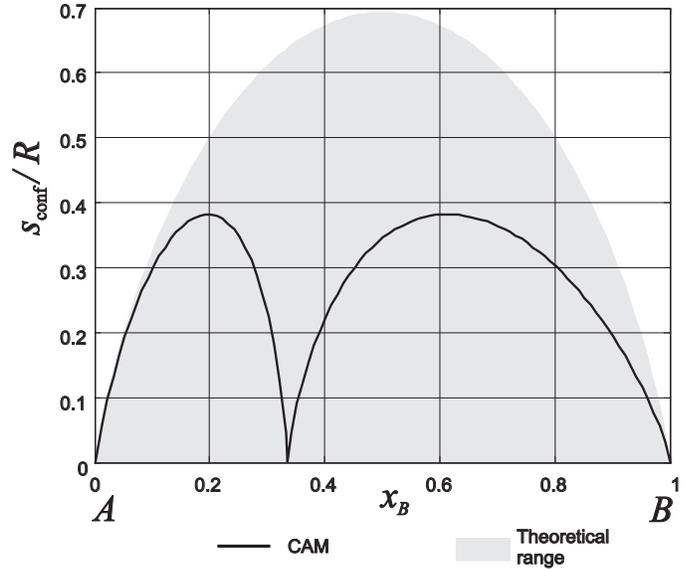}%
\caption {Molar configurational entropy of mixing given by the classical associate model
in the limit of the highly ordered solution.}
\label{fig cam ordered}%
\end{center}
\end{figure}

Now consider the limit of highly ordered solution and assume that the Gibbs free energy
change of forming $A_2B$-associates approaches $-\infty$. The entropy curve given by the
classical associate model in this case is shown in Fig.~\ref{fig cam ordered}. As shown
in this figure, the entire entropy curve is within the theoretically possible range. Note
also that the configurational entropy of mixing equals to zero at the composition of
maximum ordering, since different spatial arrangements of particles in an
$A_2B$-associate are not considered in the classical associate model.
%

\subsection{Associate Species Model}
According to the associate species model~\cite{Besmann02}, the configurational entropy of
mixing of $x_A$ moles of the component $A$ and $x_B=1-x_A$ moles of the component $B$ is
given by
\begin{equation}\label{eq associated S_config}
s_\text{conf}=-\frac{R}{\p}
    \left(
    \sum\limits_{a+b=1}n_{A_{a\cdot \p}B_{b\cdot\p}}\ln(x_{A_{a\cdot \p}B_{b\cdot \p}})
    \right).
\end{equation}
Here, $\p$ is the size of species (the number of particles per an associate specie);
$n_{A_{a\cdot \p}B_{b\cdot \p}}$ are the mole numbers and $x_{A_{a\cdot \p}B_{b\cdot
\p}}$ are the molar fractions of the ${A_{a\cdot \p}B_{b\cdot \p}}$-species. Similar to
the classical associate model, the size and compositions of species are determined by a
modeller.

\begin{figure}
\begin{center}
\includegraphics[width=\columnwidth]{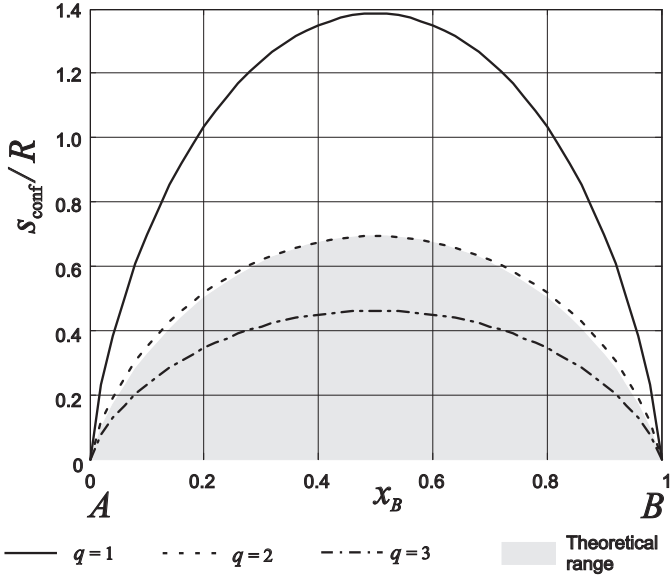}%
\caption {Molar configurational entropy of mixing given by the associate species model
with the species $A_\p$, $B_\p$, $A_{2\p/3}B_{\p/3}$ and $A_{\p/3}B_{2\p/3}$ in the limit
of ideal solution.}
\label{fig as_sp_ideal}%
\end{center}
\end{figure}

Consider the limit of ideal solution, assuming that the solution consists of the
associate species $A_\p$, $B_\p$, $A_{2\p/3}B_{\p/3}$ and $A_{\p/3}B_{2\p/3}$. The
Fig.~\ref{fig as_sp_ideal} shows the entropy curves given by the associate species model
for $\p=1$, $\p=2$ and $\p=3$. In contrast to the classical associate model, one can
adjust the configurational entropy of mixing, using the additional parameter $\p$. For
this particular set of species, the selection of $\p=2$ result in the entropy curve which
is closed (though located outside the theoretically possible range) to the entropy curve
given by the ideal solution model. As shown in Fig. ~\ref{fig as_sp_ideal}, the selection
$\p=1$ results in substantial overestimation of the configurational entropy of mixing,
while the selection $\p=3$ gives underestimated values.

\begin{figure}[!h]
\begin{center}
\includegraphics[width=\columnwidth]{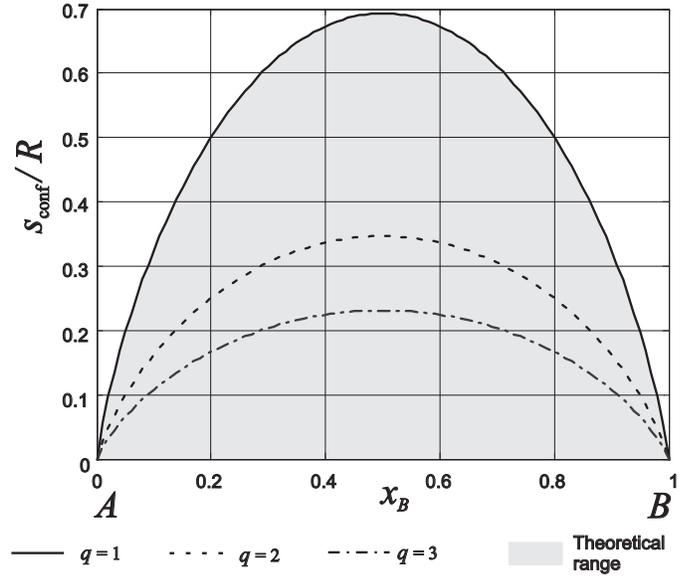}%
\caption {Molar configurational entropy of mixing given by the associate species model in
the limit of immiscible components.}
\label{fig as_sp_imm}%
\end{center}
\end{figure}

Consider the limit of immiscible components, that is, the Gibbs free energy changes on
forming the species $A_{2\p/3}B_{\p/3}$ and $A_{\p/3}B_{2\p/3}$ approach $+\infty$. The
entropy curves given by the associate species model in this case are shown in
Fig.~\ref{fig as_sp_imm}. As shown in this figure, the associate species model
overestimate the configurational entropy of mixing in the limit of immiscible components.
However, the entropy curve tends to theoretically expected one ($s_\text{conf}\equiv0$),
when $\p$ increases.

\begin{figure}[!h]
\begin{center}
\includegraphics[width=\columnwidth]{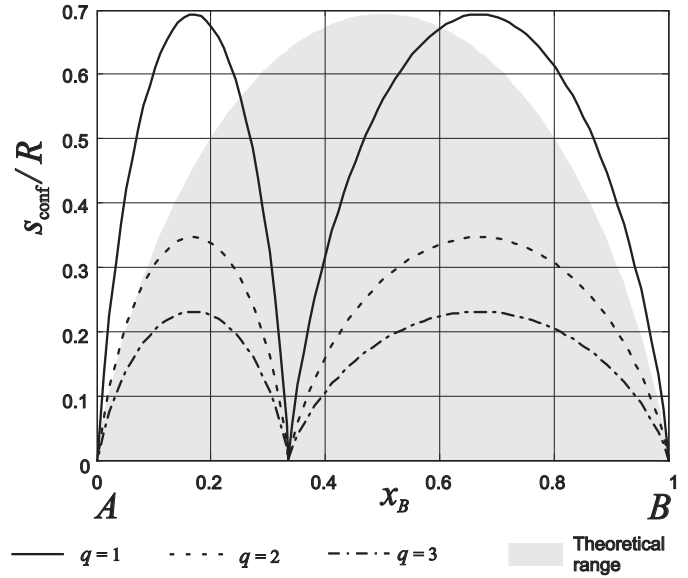}%
\caption {Molar configurational entropy of mixing given by the associate species model in
the limit of highly ordered solution.}
\label{fig as_sp_ord}%
\end{center}
\end{figure}

Now consider the limit of highly ordered solution. Assume also that the Gibbs free energy
change on forming the associate species $A_{2\p/3}B_{\p/3}$ approaches $-\infty$.
Fig.~\ref{fig as_sp_ord} shows the entropy curve given by the associate species model in
this case. As seen from Fig.~\ref{fig as_sp_ord}, the entropy curve with $\p=3$ is
entirely located in the theoretically possible range. If $\p=2$, however, small part of
the entropy curve (approximately, for $0<x_B<0.075$) is located outside the theoretically
possible range. The selection $\p=1$ results in paradoxical values of configurational
entropy approximately for $0<x_B<0.25$ and $0.6<x_B<1$.
%

\subsection{Modified Associate Formalism}
In the modified associate formalism the configurational entropy of mixing is given by
(see Eq.~(44) in Ref.~\cite{Saulov_MAF1})
\begin{equation}\label{eq maf s_conf}
s_\text{conf}= -R \sum\limits_{\kk \in\set}
        \sum\limits_{j=1}^{J_\kk}
        n_\kk^\jj
        \ln\left(
        \frac{x_\kk^\jj}{d_\kk^\jj}\right)~,
\end{equation}
where all the notations have the same meaning as in Ref.~\cite{Saulov_MAF1}. In contrast
to previous modifications of the associate model, the compositions of associates that
have to be considered are defined by the size of associates.

Consider the limit of ideal solution. In this limit, the Gibbs free energies of the
reactions of forming different associates equal to zero. As demonstrated in the previous
paper~\cite{Saulov_MAF1}, the configurational entropy of mixing correctly reduces to that
of the ideal solution model for arbitrary size of associates.

\begin{figure}
\begin{center}
\includegraphics[width=\columnwidth]{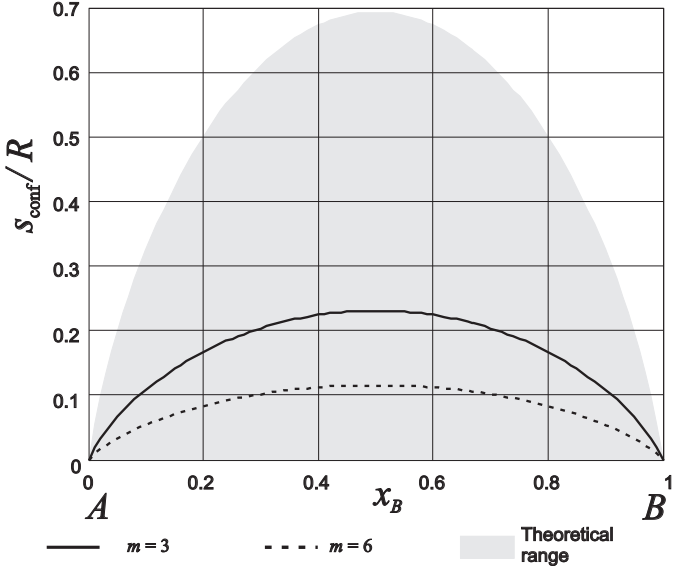}%
\caption {Molar configurational entropy of mixing given by the modified associate
formalism in the limit of immiscible components.}
\label{fig maf_imm}%
\end{center}
\end{figure}

Now consider the limit of immiscible components, where the Gibbs free energies of the
associate that consist of both $A$-particles and $B$-particles approach $+\infty$. In
this case, Eq.~(\ref{eq maf s_conf}) reduces to
\begin{equation}\label{eq associated S_config}
s_\text{conf}=-\frac{R}{m}(x_{A}\ln(x_{A})+n_{B}\ln(x_{B})).
\end{equation}
where $m$ is the size of associates. The entropy curves for $m=3$ and $m=6$ are shown in
Fig.~\ref{fig maf_imm}. As seen from this figure, the entropy curves are entirely located
in the theoretically possible range and tends to the theoretically expected curve
($s_\text{conf}\equiv 0$) with increasing the size of associates.

\begin{figure}
\begin{center}
\includegraphics[width=\columnwidth]{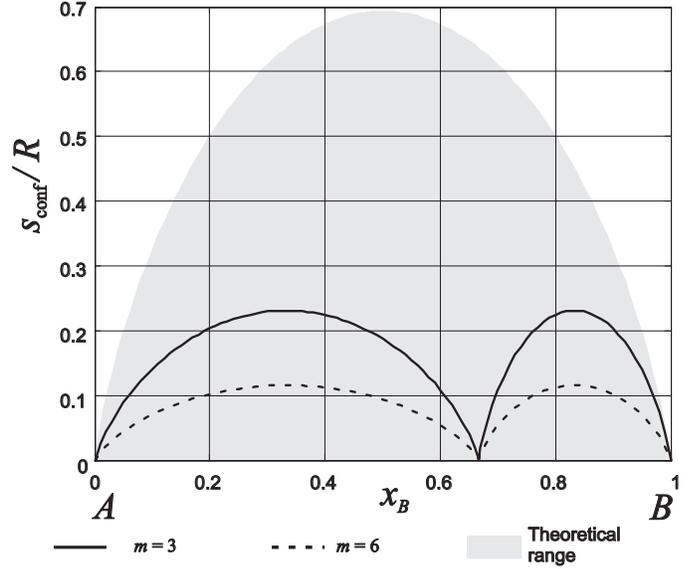}%
\caption {Molar configurational entropy of mixing given by the modified associate
formalism in the limit of highly ordered solution.}
\label{fig maf_ord}%
\end{center}
\end{figure}

Considering the limit of highly ordered solution, we assume also that the Gibbs free
energy of associate with the composition $A_2B$ and with a particular spatial arrangement
of particles approaches $-\infty$. The entropy curves for $m=3$ and $m=6$ given by the
modified associate formalism in this case is presented in Fig.~\ref{fig maf_ord}. In this
case, the curves are also entirely located in the theoretically possible range.
%

\subsection{Modified Quasichemical Model (1986)}
In the modified quasichemical model, the molar configurational entropy of mixing is given
by (see Eq.~(10) in Ref.~\cite{Pelton_00})
\begin{equation}\label{eq qc s_conf}
\begin{array}{l}
s_\text{conf}=-R(x_A \ln (x_A)+x_B \ln (x_B))\\
- R\left[n_{AA} \ln \left(\frac {x_{AA} }{y_A^{2} } \right)+n_{BB} \ln \left(  \frac
{x_{BB} } {y_B^{2} } \right)+ n_{AB}\ln \left( \frac {x_{AB}} {2y_A y_B } \right)\right].
\end{array}
\end{equation}
Here, $x_A$ and $x_B$ are the molar fractions of the components $A$ and  $B$; $n_{ij}$
and $x_{ij}$ ($i,j = A,B$) are the mole number and the molar fraction of $(i-j)$-pairs,
respecively. The coordination-equivalent fractions $y_A$ and $y_B$ are defined as (see
Eq.~(6) in Ref.~\cite{Pelton_00})
\begin{equation}
\begin{array}{ll}
y_{A}&={Z_{A}n_A}/({Z_{A}n_A+Z_{B}n_B})\\
y_{B}&={Z_{B}n_B}/({Z_{A}n_A+Z_{B}n_B})~,
\end{array}
\end{equation}
where $Z_A$ and $Z_B$ are the coordination numbers of the solution components  $A$ and
$B$, respectively. As demonstrated by Pelton and Blander~\cite{Pelton_86} for constant
coordination numbers and by Pelton et al.~\cite{Pelton_00} for variable coordination
numbers, the modified quasichemical model correctly reduces to the ideal solution model,
when the Gibbs free energy change on forming $(A-B)$-pairs $\Delta g_{AB}$ equals to
zero.

In the framework of MQM(1986) (the modified quasichemical model after Pelton and
Blander~\cite{Pelton_86}), the coordination numbers are assumed to be constant and are
selected to set the composition of maximum ordering $x_B^*$ and to satisfy the conditions
$s_\text{conf}=0$ at $x_B^*$ when $\Delta g_{AB}\rightarrow-\infty$. This results in (see
also Eqs. (30) and (31) in Ref.~\cite{Pelton_86})
\begin{equation}\label{eq qc s_conf}
\begin{array}{ll}
Z_B&=-(x_B^*\ln(x_B^*)+(1-x_B^*)\ln(1-x_B^*))/(x_B^*\ln(2))\\
Z_A&=Z_Bx_B^*/(1-x_B^*).
\end{array}
\end{equation}

\begin{figure}
\begin{center}
\includegraphics[width=\columnwidth]{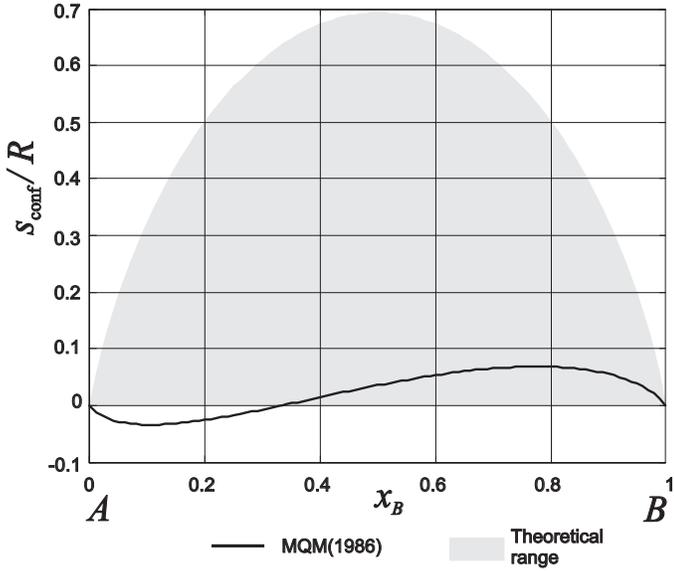}%
\caption {Molar configurational entropy of mixing given by the modified quasichemical
model (1986) in the limit of immiscible components.}
\label{fig mqc1_imm}%
\end{center}
\end{figure}

Similar to the examples described above, consider one mole of binary solution $A-B$ and
assume that $x_B^*=1/3$. In this case, $Z_B=2.75489$ and $Z_A=1.37744$. In the limit of
immiscible components, $\Delta g_{AB}\rightarrow+\infty$. The molar configurational
entropy of mixing given by the curves given by MQM(1986) in this case is presented in
Fig.~\ref{fig mqc1_imm}. As shown in this figure, MQM(1986) gives paradoxical, negative
values of the configurational entropy of mixing for $0<x_B<1/3$.

\begin{figure}[!h]
\begin{center}
\includegraphics[width=\columnwidth]{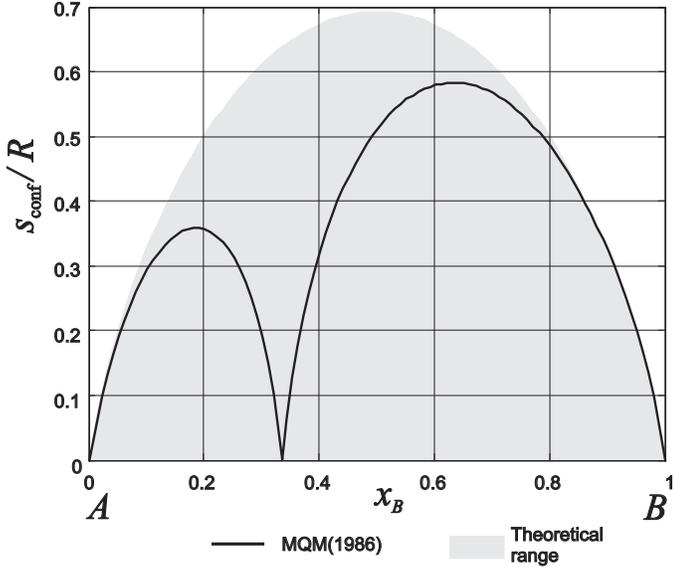}%
\caption {Molar configurational entropy of mixing given by the modified quasichemical
model (1986) in the limit of highly ordered solution.}
\label{fig mqc1_ord}%
\end{center}
\end{figure}

Now consider the limit of highly ordered liquid, that is, $\Delta
g_{AB}\rightarrow-\infty$. Fig.~\ref{fig mqc1_ord} shows the entropy curve given by
MQM(1986) in this case. As seen from Fig.~\ref{fig mqc1_ord}, the curve is entirely
located in the theoretically possible range.

\subsection{Modified Quasichemical Model (2000)}
In the framework of MQM(2000) (the modified quasichemical model after Pelton et
al.~\cite{Pelton_00}), the configurational numbers are assumed to be
composition-dependant. The following equations for the coordination numbers have been
suggested (see Eqs. (19) and (20) in Ref. \cite{Pelton_00}):
\begin{equation}\label{eq variable coordination numbers}
\begin{array}{ll}
\frac {1}{Z_A}&= \frac {1}{Z^A_{AA}} \frac {2n_{AA}}{2n_{AA}+n_{AB}}+
\frac {1}{Z^A_{AB}} \frac {n_{AB}}{2n_{AA}+n_{AB}}\vspace{.2 cm} \\
\frac {1}{Z_B}&= \frac {1}{Z^B_{BB}} \frac {2n_{BB}}{2n_{BB}+n_{AB}}+ \frac {1}{Z^B_{BA}}
\frac {n_{AB}}{2n_{BB}+n_{AB}}~.
\end{array}
\end{equation}

\begin{figure}[!h]
\begin{center}
\includegraphics[width=\columnwidth]{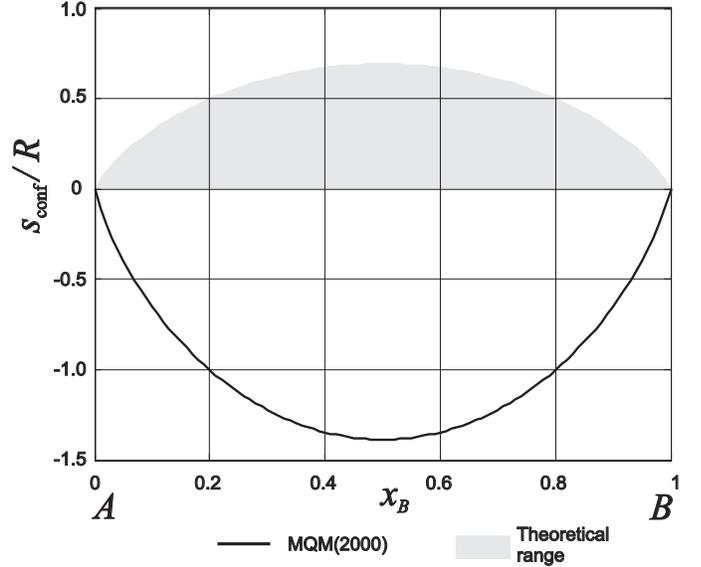}%
\caption {Molar configurational entropy of mixing given by the modified quasichemical
model (2000) in the limit of immiscible components.}
\label{fig mqc2_imm}%
\end{center}
\end{figure}

Consider the limit of immiscible components ($\Delta g_{AB}\rightarrow+\infty$), assuming
that the constants $Z^A_{AA}$, $Z^B_{BB}$, $Z^A_{AB}$ and $Z^B_{BA}$ are chosen as
suggested by Pelton et al.~\cite{Pelton_00}, namely:
\begin{equation}\label{ZZZ}
Z^A_{AA}=Z^B_{BB}=Z^B_{BA}=6;~~Z^A_{AB}=3~.
\end{equation}
Fig.~\ref{fig mqc2_imm} demonstrates that MQM(2000) gives paradoxical values of the
configurational entropy of mixing for the entire compositional range in the limit of of
immiscible components.

\begin{figure}[!h]
\begin{center}
\includegraphics[width=\columnwidth]{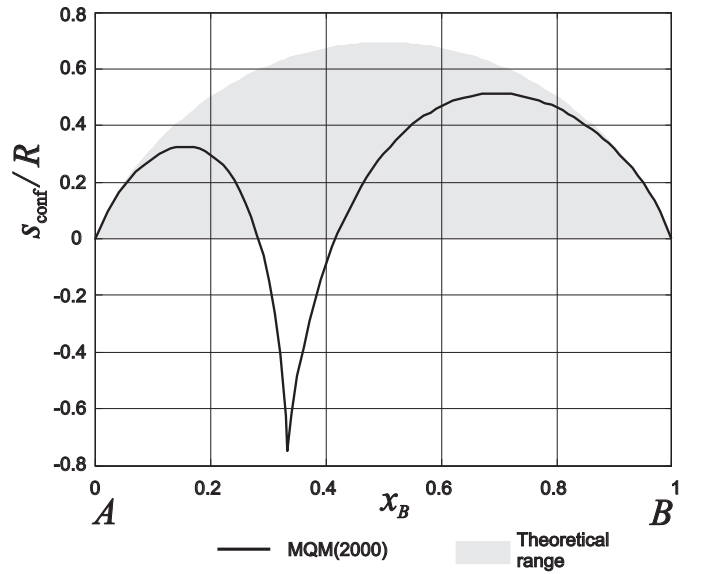}%
\caption {Molar configurational entropy of mixing given by the modified quasichemical
model (2000) in the limit of highly ordered solution.}
\label{fig mqc2_ord}%
\end{center}
\end{figure}

Now consider the limit of highly ordered solution, that is, $\Delta
g_{AB}\rightarrow-\infty$. The entropy curve given by MQM(2000) in this case is presented
in Fig.~\ref{fig mqc2_ord}. As seen form this figure, MQM(2000) gives negative values of
the configurational entropy of mixing in the vicinity of the composition of maximum
ordering.
%

\subsection{Classical Quasichemical Model}
It is important to note that the classical quasichemical model of Fowler and
Guggenheim~\cite{Fowler_Guggenheim} (with $Z_A=Z_B=1/2$) correctly reduces to the ideal
solution model when $\Delta g_{AB}=0$. In the limit of immiscible components, the
configurational entropy of mixing given by this model correctly reduces to the
theoretically expected curve ($s_\text{conf}\equiv 0$). In all other cases, the classical
quasichemical model gives values of the configurational entropy of mixing which are
within the theoretically possible range. The applicability of the classical quasichemical
model, however, is limited to those chemical solutions, in which maximum ordering occurs
at the composition $x_B=1/2$.

\subsection{Summary}
The classical associate model results in paradoxically large values of the
configurational entropy in the limit of ideal solution, while giving theoretically
possible values of the configurational entropy of mixing in other two limits. The
associate species model can give paradoxically large values in the limit of ideal
solution and in the limit of highly ordered solution.

The modified quasichemical model after Pelton and Blander~\cite{Pelton_86} results in
paradoxically small, negative values of the configurational entropy in the limit of
immiscible components. The modified quasichemical model after Pelton et
al.~\cite{Pelton_00} gives paradoxically small, negative values in the limit of
immiscible components and in the limit of highly ordered solution.

The modified associate formalism~\cite{Saulov_MAF1} and the classical quasichemical
model~\cite{Fowler_Guggenheim} give theoretically possible values of the configurational
entropy in all considered limits. The classical quasichemical model also correctly
reduces to the theoretically expected curve in the limit of immiscible components. The
applicability of the classical quasichemical model, however, is limited.

\section{Dilute Solutions}\label{sec dilute}
Pelton et al. \cite{Pelton_00} considered binary solution, in which maximum ordering
occurs at the composition $x_B=1/3$ in the limit of highly ordered solution (Gibbs free
energy change on forming $(A-B)$-pairs approaches $-\infty$). It was demonstrated that,
in the framework of the modified quasichemical model, chemical activity $a_{B}$ of the
component $B$ in the varies in the vicinity of $x_{A}=0$ as ($1-x_{A}$) even in this
limiting case.

Similar solution has also been considered in Ref.~\cite{Saulov_MAF1} in the framework of
the modified associate formalism. As demonstrated, chemical activity $a_{B}$ varies as
($1-x_{A}$) only for finite values of $\Delta g_{2,1}$ and $\Delta g_{1,2}$. Here,
$\Delta g_{2,1}$ and $\Delta g_{1,2}$ are the molar Gibbs free energy changes on forming
$A_2B$-associates and $AB_2$-associates, respectively. In the limit of highly ordered
solution ($\Delta g_{2,1}\rightarrow-\infty$), $a_{B}$ varies as ($1-x_{A}/2$). Note also
that the interval in the vicinity of $x_{A}=0$, in which $a_{B}$ can be approximated as
($1-x_{A}$), decreases with the decrease in $\Delta g_{2,1}$. This should be taken into
consideration, when the modified associate formalism is used for modelling dilute
solutions.

Now consider the solution discussed in in Ref.~\cite{Saulov_MAF1} in the framework of the
associate species model. That is, we consider the associate species $A_\p$, $B_\p$,
$A_{2\p/3}B_{\p/3}$ and $A_{\p/3}B_{2\p/3}$ of the size $\p$. In this case, chemical
activity $a_{B}$ is given by
\begin{equation}\label{activity associate species}
a_B=x_{B_\p}^{1/ \p}~,
\end{equation}
where $x_{B_\p}$ is the molar fraction of the species $B_\p$. Eq.~(28) in
Ref.~\cite{Saulov_MAF1} reads
\begin{equation}\label{eq da}
\frac{da_B}{dx_A}=\frac{1}{\p}x_{B_\p}^{1/\p-1}\left(\frac{dx_A}{dx_{B_\p}}\right)^{-1}~.%
\end{equation}
Using the technique described in Section 6 of the previous paper~\cite{Saulov_MAF1}, one
verifies that
\begin{equation}\label{eq mb4}
\begin{array}{ll}
\frac{dx_{A}}{dx_{B_\p}}&=-1-\frac{4}{9}\epsilon_{2} t -\frac{1}{9}\epsilon_{1} t^2\\
&+\frac{2}{9}\frac{\left(\epsilon_{2} +\epsilon_{1}
t\right)\left(1+\frac{2}{3}\epsilon_{2} t+ \frac{1}{3}\epsilon_{1} t^{2}
\right)}{\left(t^{2}+\frac{1}{3}\epsilon_{2} + \frac{2}{3}\epsilon_{1} t \right)}
\end{array}
\end{equation}
Here, $t \equiv \left({x_{A_\p}}/{x_{B_\p}}\right)^{1/3}$, $\epsilon_{1} \equiv
\exp\left(-{\Delta g_{A_{2\p/3}B_{\p/3}}}/{RT}\right)$ and $\epsilon_{2} \equiv $
$\exp\left(-{\Delta g_{A_{\p/3}B_{2\p/3}}}/{RT}\right)$, while $\Delta
g_{A_{2\p/3}B_{\p/3}}$ and $\Delta g_{A_{\p/3}B_{2\p/3}}$ are the molar Gibbs free
energies of formation of $A_{2\p/3}B_{\p/3}$-species and $A_{\p/3}B_{2\p/3}$-species,
respecively.

Substituting Eq.~(\ref{eq mb4}) into Eq.~(\ref{eq da}) and taking the limit when
$x_{A}\rightarrow 0$, one verifies that
\begin{equation}
\lim \limits_{x_{A}\rightarrow 0}(da_B/dx_A)=-\frac{3}{\p}~.
\end{equation}
This demonstrates that, for any finite values of $\Delta g_{A_{2\p/3}B_{\p/3}}$ and
$\Delta g_{A_{\p/3}B_{2\p/3}}$, chemical activity $a_B$ given by the associate species
model varies as ($1-x_{A}$) only when $\p=3$. In the limit of highly ordered solution
($\Delta g_{A_{2\p/3}B_{\p/3}}\rightarrow-\infty$), $a_B$ varies as
($1-\frac{3}{2\p}x_{A}$) in the vicinity of $x_{A}=0$.

It is also important to note that, in the framework of the associate species model, one
can consider the species $A_\p$, $B_\p$ and $A_{2\p/3}B_{\p/3}$ only, since the set of
species to be considered is determined by a modeller. Formally setting $\Delta
g_{A_{\p/3}B_{2\p/3}}=+\infty$, Eq.~(\ref{eq mb4}) reduces to
\begin{equation}\label{eq mb5}
\frac{dx_{A}}{dx_{B_\p}}=-1 -\frac{1}{9}\epsilon_{1} t^2 +\frac{2}{9}\frac{\epsilon_{1} t
\left(1+\frac{1}{3}\epsilon_{1} t^{2} \right)}{\left(t^{2}+\frac{2}{3}\epsilon_{1} t
\right)}~.
\end{equation}
Substitution of Eq.~(\ref{eq mb5}) into Eq.~(\ref{eq da}) and taking the limit when
$x_{A}\rightarrow 0$ gives
\begin{equation}
\lim \limits_{x_{A}\rightarrow 0}(da_B/dx_A)=-\frac{3}{2\p}~.
\end{equation}
In this case, $a_B$ varies in the vicinity of $x_{A}=0$ as ($1-\frac{3}{2\p}x_{A}$) for
\emph{any} value of $\Delta g_{A_{2\p/3}B_{\p/3}}$.

The examples presented above indicate that the set of associate species and the size of
species should be carefully selected, when the associate species model is applied for
dilute solutions.

\section{Setting the Composition of Maximum Ordering}
Modelling of multicomponent systems requires a provision for setting the composition of
maximum ordering so as to comply with the experimental data for each sub-system (binary,
ternary, quaternary and so on) individually. Associate-type models allow to do this in a
straightforward way by including in each sub-system the associates, which composition
coincides with that of maximum ordering in a particular sub-system. In the case of the
classical associate model or the associate species model, a modeller can arbitrarily
include the required associates or the associate species. In the case of the modified
associate formalism, in which the compositions of associates are defined by their size,
one should select the adequately large size of associates to include the required
compositions.

In contrast to associate-type models, there are no provisions for setting the
composition of maximum ordering for each ternary, quaternary and so on
sub-systems individually neither in the framework of MQM(1986) nor in the
framework of MQM(2000). Furthermore, the composition of maximum ordering can
not be set independently even for binary sub-systems in the framework of
MQM(1986). To overcome this disadvantage for binary sub-systems, Pelton et
al.~\cite{Pelton_00} suggested to use variable coordination numbers given by
Eqs.~(\ref{eq variable coordination numbers}). This modification, however,
leads to mathematical and thermodynamical inconsistencies of MQM(2000). These
inconsistencies are described immediately below.

Consider the case, in which the coordination numbers are given by Eqs.~(\ref{eq variable
coordination numbers}) and the constants $Z^A_{AA}$, $Z^B_{BB}$, $Z^A_{AB}$ and
$Z^B_{BA}$ are given by Eqs. (\ref{ZZZ}). Firstly, as demonstrated in
Ref.~\cite{Saulov_QCM_modifications}, the total mole number of pairs $n_\text{tp}$
depends on the extent of the quasichemical reaction (Eq.~(1) in Ref.~\cite{Pelton_00}).
More precisely, $n_\text{tp}$ decreases by $1/2$ mole per each mole of $(A-B)$ pairs
formed. This result contradicts to the equation of the quasichemical reaction (see
Eq.~(1) in Ref.~\cite{Pelton_00}), which underlies the quasichemical model. According to
Eq.~(1) in Ref.~\cite{Pelton_00}, $n_\text{tp}$ does not change in the quasichemical
reaction. Secondly, when the coordination numbers $Z_A$ and $Z_B$ vary with $n_{AB}$, the
expression for the Gibbs free energy of the solution given by Eq.~(9) in
Ref.~\cite{Pelton_00}) is in error. Using Eqs.~(1-3,19,20) in Ref.~\cite{Pelton_00}, one
verifies that the correct expression in this case is
\begin{equation}\label{}
G=\frac{Z_A}{Z^A_{AA}} n_A g_A^\circ+ \frac{Z_B}{Z^B_{BB}} n_B g_B^\circ %
-T\Delta S^\text{config}+ \frac{n_{AB}}{2} \Delta g_{AB}~.
\end{equation}
Here, all the notations have the same meaning as in Ref.~\cite{Pelton_00}.

\section{Conclusions}
The modified associate formalism has been compared with similar models. As demonstrated,
the formalism gives theoretically possible values of the configurational entropy in all
considered limits. Being an associate-type model, the formalism allows to set the
composition of maximum ordering for each sub-system of a multicomponent system
individually. The corrected Gibbs free energy equation of the quasichemical model is also
presented.

\section*{Acknowledgment}
The present work has been supported by the Australian Research Council.

\bibliographystyle{elsart-num}


\end{document}